\documentclass[aps,showpacs,twocolumn,pre]{revtex4}

\usepackage{graphicx}
\usepackage{amssymb}
\usepackage{natbib}
\usepackage{amsmath}

\begin{document}
\title{Crystallization and gelation in colloidal systems with short-ranged attractive interactions}
\author{Andrea Fortini\footnote{Present address:
Department of Physics, Yeshiva University, 2495 Amsterdam Avenue, New York, NY 10033, USA}}
\author{Eduardo Sanz\footnote{Present address:
SUPA School of Physics,
JCMB The Kings Buildings,
University of Edinburgh,
Mayfield Road
Edinburgh EH9 3JZ, UK}}
\author{Marjolein Dijkstra}
\email{m.dijkstra1@uu.nl}
\affiliation{ Debye Institute for NanoMaterials Science, Utrecht
University, \\ \small Princetonplein 5, 3584 CC Utrecht, The 
Netherlands }

\begin{abstract}
We systematically study the relationship between equilibrium and non-equilibrium phase diagrams of a system of short-ranged attractive colloids. 
Using Monte Carlo and
Brownian dynamics simulations we find a
window of enhanced crystallization that is limited at high interaction strength by a slowing down of the dynamics and at low interaction strength by the high nucleation barrier. 
We find that the crystallization is
enhanced by the metastable gas-liquid binodal by means of a
two-stage crystallization process. First, the formation of a dense
liquid is observed and second the crystal nucleates within the
dense fluid. 
In addition, we find at low colloid packing fractions 
a fluid of clusters, and at higher colloid packing
fractions a percolating network due to an arrested gas-liquid
phase separation that we identify with gelation. 
We find that this arrest is due to crystallization at low interaction energy and it
is caused by a slowing down of the dynamics at high interaction strength. Likewise,
we observe that the clusters which are formed at low colloid
packing fractions are crystalline at low interaction energy, but
glassy at high interaction energy. The clusters coalesce upon encounter. 
\end{abstract}


\maketitle

\section{Introduction}
Crystallization, vitrification and gelation in systems with
short-ranged attractive interactions like colloidal particles and
proteins are widely studied
phenomena~\cite{Bos1996,Verhaegh1999,
Hoog2001,Anderson2001,Bergenholtz2003,Sedgwick2005,Manley2005,Zaccarelli2007,Bailey:2007qf}.
Crystallization kinetics is especially important for proteins
because good crystals are necessary for the determination of their
3D structure. Furthermore, crystallization of biologically active
proteins can lead to diseases like the human genetic cataract~\cite{Pande:2001}.
While the crystal is an equilibrium
phase~\cite{Gast1983,Meijer1991,Lekkerkerker1992,Meijer1994,Dijkstra1999a}, gels and glasses
are non-equilibrium phases that occur
due to dynamical arrest~\cite{Pusey1986,Weeks2000,Kegel2000,Pham2002,Anderson2002}. 
Vitrification in a fluid of $N_c$ hard
spheres with diameter $\sigma_c$ in a volume $V$, occurs for
packing fractions $\eta =(\pi \sigma_c^3/6)N_c/V \gtrsim
0.58$~\cite{Pusey1986}, while the equilibrium phase diagram
predicts a stable crystal phase for $\eta>0.545$~\cite{Hoover1968}. 
In the \emph{repulsive} colloidal glass phase, the particles are arrested as
they become localized in a ``cage'' formed by the surrounding
spheres. Interestingly, the addition of
non-adsorbing polymer, resulting in an effective attraction
between the colloids, can enhance the relaxation process so that the system can crystallize.
Nevertheless, for increasing attraction strengths the particles
become more and more localized due to long-lived bonds between the
neighboring particles until the system is arrested in an
\emph{attractive} glass at smaller packing fractions~\cite{Pham2002}. 
An arrested state at low $\eta$ is often called a gel. 
However, the relation between gels and attractive glasses is still highly debated. In general,
gels can be classified as chemical, if the particles are linked by
irreversible bonds, or as physical, if the bonds are not
permanent. Physical gels can form through a non-equilibrium
mechanism  due to the arrest of a (metastable) gas-liquid
separation by the attractive glass transition~\cite{Lu:2008,Sanz:2008}. On the other hand,
equilibrium gelation can occur in systems  where the gas-liquid
phase separation is suppressed by e.g. long-ranged repulsions or
directional interactions~\cite{Zaccarelli2007}. For increasing
contact energy these two mechanisms for the formation of physical
gels continuously evolve towards a diffusion limited cluster
aggregation (DLCA) that leads to chemical gels. The metastable
gas-liquid separation is therefore important for the formation of
non-equilibrium gels~\cite{Lu:2008}, and it is widely recognized as an important
factor in protein crystallization~\cite{Tanaka1997, Lomakin2003,
Filobelo2005, Lutsko2006}.  There has been much recent experimental, theoretical and simulation effort in understanding 
the equilibrium phase diagram or the kinetics and dynamics of these systems, but only a few studies explored the relationship between 
the equilibrium phase diagram and the underlying kinetic pathways for
gelation and crystallization. Compared to previous
works~\cite{Foffi2005,Charbonneau2007} we systematically
determine the crystallization behavior in relation to the
equilibrium phase diagram in the region of the metastable
gas-liquid phase separation and we investigate the relation
between crystallization and gelation.

\section{Model and Method}
We study a model system of $N_{c}$ hard spheres of diameter
$\sigma_{c}$ and non-adsorbing polymers with diameter $\sigma_{p}$
described by the Asakura-Oosawa-Vrij (AOV) model~\cite{Vrij1976}. 
The polymers induce an effective
short-ranged attractive depletion interaction between the colloids~\cite{Asakura1954,Vrij1976}.
The total interaction potential is defined as 
\begin{equation} 
\beta U_{\rm tot}({r}_{ij})=\left \{ 
\begin{array}{ll}
 \infty  &  r_{ij}<  \sigma_c\\
  \beta U_{\rm dep}(r_{ij})
    &  \sigma_c<r_{ij}<  \sigma_c+\sigma_p \\
  0 & r_{ij}> \sigma_{c}+\sigma_p 
\end{array} \right . \ ,
\label{te:tot}
\end{equation}
with
\begin{eqnarray}
\lefteqn{\beta U_{\rm dep}({r}_{ij})=} \nonumber \\
& -\eta_{p}^{r} \frac{(1+q)^3}{q^3}\left [ 1- \frac{3 r_{ij}}{2
(1+q)\sigma_c} + \frac{r_{ij}^3}{2(1+q)^3\sigma_c^3}\right ] \ ,
\label{te:aopot1}
\end{eqnarray}
where
$r_{ij}=|\vec{R}_i-\vec{R}_j|$ is the distance between two
colloidal particles with $\vec{R}_i$ the center-of-mass of colloid
$i$, and $\beta=1/k_{B} T$ with $k_{B}$ the Boltzmann's constant and
$T$ the temperature.  Thus the size ratio $q=\sigma_{p}/\sigma_{c}$
determines the range of the interaction. The polymer reservoir
packing fraction $\eta_{p}^{r}=(\pi \sigma_{p}^{3}/6)
\rho_{p}^{r}$, is proportional to the polymer density
$\rho_{p}^{r}$ in the reservoir with which our system is in
osmotic contact. This parameter controls the strength of
attraction. As a measure of this interaction
strength we also use the contact energy $\beta U=-\beta U_{\rm
dep}({r}_{ij}=\sigma_{c})$ and the experimentally accessible reduced second virial coefficient $B_{2}^{*}=1+3/\sigma_{c}^{3} \int_{\sigma_{c}}^{\sigma_{c}+\sigma_{p}} r^{2}(1-\exp(-\beta U_{\rm dep}(r)))dr$, which is normalized by the second virial coefficient of hard spheres with a diameter $\sigma_{c}$.
Although colloids often have a Coulombic  repulsion~\cite{Sciortino2005} 
and the attractive potential in proteins is
often patchy~\cite{Piazza2000} our simple model captures the
basic phenomena that play a role in the competition between
equilibrium and non-equilibrium phases of systems with
short-ranged attractions. 

We determine the equilibrium phase diagram by calculating the
dimensionless free energy density $f= \beta F /V$ as a
function of the colloid packing fraction $\eta_c$ and contact energy $\beta U$ (polymer reservoir packing fraction  $\eta_{p}^{r}$) with Monte Carlo simulations. 
We use thermodynamic integration to relate the 
free energy of the effective system to that of a reference system
at the same colloid volume fraction $\eta_c$.
To this end, we write the total free energy density as the sum of
two contributions
\begin{equation}
f(N_c,V,\eta_{p}^{r}) = f_{c}(N_c,V) +
f_{\rm dep}(N_c,V,\eta_{p}^{r}) \ ,
 \label{E:free-en}
\end{equation}
where  $f_{c}(N_c,V)$ is the free energy density of a system of
$N_c$ hard spheres in a volume $V$, for which we used the Carnahan-Starling expression~\cite{Carnahan1969} for the fluid phase, and the Hall expression~\cite{Hall1972} for the face-centred-cubic (f.c.c.) solid phase.
The free energy density $f_{\rm dep}$ is the contribution of the
depletion potential (\ref{te:aopot1}) to the free energy density, and it is computed using the $\lambda$-integration~\cite{Frenkel2002}.
In order to map out the phase diagram we determine the total free energy density
$f(\eta_{c},\eta_p^{r})$ for many state points $(\eta_{c},\eta_p^{r})$ in
simulations, and we employ  common tangent constructions at fixed
$\eta_p^{r}$ to obtain the coexisting phases~\cite{Dijkstra1999}.
To trace the gas-solid binodal we use $N_c$=512 particles with $15\times10^{4}$  Monte Carlo moves per particle for equilibration and $15\times10^{4}$  Monte Carlo moves per particle for production. To check the accuracy of the results few simulations were repeated using $15\times10^{5}$  Monte Carlo moves per particle for equilibration and $15\times10^{5}$  Monte Carlo moves per particle for production. 
To trace the metastable binodal while avoiding crystallization~\cite{Dijkstra1999,Fortini2006b,Caballero:2006} we use a smaller system with $N_c$=120 particles. 
To get good statistics 5 to 10 independent Monte Carlo runs are necessary for each value of the density.

To study the dynamics of the system we perform standard Brownian
dynamics (BD) simulations~\cite{Allen1987} by  quenching
homogeneous configurations to one of the state points indicated by
the points in Fig.~\ref{fig:phdfull2}. The time step $\delta t$
is chosen to be $\delta t =2 \times 10^{-5} \tau_{B}$ with
$\tau_B=\sigma_c^2/(4 D_0)$, where $D_{0}$ is the Stokes-Einstein
diffusion coefficient. We neglect all hydrodynamic interactions
between the particles. The random forces  mimic the interaction
between particles and solvent, and are sampled from a Gaussian
distribution with variance $2 D_{0} \delta t$. The random force
and the dissipative term provide the system with a heat bath at
constant temperature. We carry out BD simulations  using particle
numbers $N_{c}$ ranging from 600 to 1200 and to avoid discontinuities in the potential we replace the hard-core
potential with the equivalent repulsive soft potential $1/r^{36}$. We studied
the evolution of a system for a total time $t=600 \tau_{B}$.
In order to study the dynamical arrest, we compute the bond
correlation function
\begin{equation}
\Phi_{B}(t)= \frac{1}{N_{c}} \sum_{i=1} \frac{\sum_{j \neq i}
n_{ij}(t) n_{ij}(0)}{\sum_{j \neq i}n_{ij}(0)}
\end{equation}
where $n_{ij}$=1 if the separation between particle $i$ and $j$ is
smaller than a  cutoff value of 1.25~$\sigma_{c}$, otherwise
$n_{ij}$=0.  The cutoff value corresponds to the minima between the first and second peak of the radial distribution function.
The function $\Phi_{B}(t)$ is similar in spirit to the one used
in~\cite{Sciortino2005}, and counts which fraction of the bonds
 at $t$=0 is still present at time $t$.

We characterise the local structure around a particle $i$ by a set of numbers
\begin{equation}
q_{lm}(i)= \frac{1}{N_{b}(i)} \sum_{j}^{N_{b}(i)} Y_{lm}(\hat r_{ij}) \ ,
\label{qlm}
\end{equation}
where $Y_{lm}(\hat r_{ij})$ are spherical harmonics, $\hat r_{ij}$ is a unit vector in the direction of the bond between particle $i$ and particle $j$.
The sum runs over all $N_{b}(i)$ neighbours of particle $i$. 
We then construct the dot product 
\begin{equation}
q(ij)=\vec q_{l}(i)  \cdot \vec q_{l}(j)= \sum_{m=-l}^{l} q_{lm}(i) q^{*}_{lm}(j)\ ,
\label{dotq}
\end{equation}
where $i$ and $j$ are neighbouring particles, and $q^{*}_{lm}(i)$ is the complex conjugate of $q_{lm}(i)$.
We normalized the vector $\vec q_{l}(i)$, such that $\vec q_{l}(i)  \cdot \vec q_{l}(i)=1$.
In our analysis particles $i$ and $j$ are defined as neighbors if $n_{ij}=1$, i.e. $R_{ij}< 1.25 \sigma_{c}$.
Two particles are defined to be joined by a crystal bond if $q(ij)>0.5$. 
Particle $i$ is  crystal-like if at least 7 of its bonds with neighbouring particles are crystal-like~\cite{Wolde1995}.   As in other studies~\cite{Gasser2001} we use the $l$=6 order parameter.

\section{Results}
\begin{figure}[htbp]
   \centering
   \includegraphics[width=8cm]{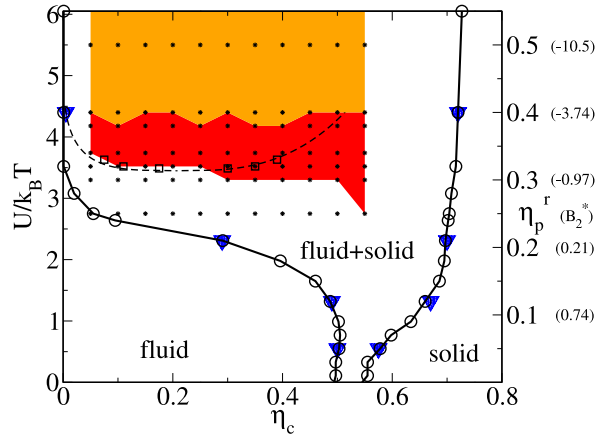}
      \caption{
     (Color online) The equilibrium phase diagram of a model colloid-polymer mixture with size ratio $q=0.15$ as a function of the colloid packing fraction $\eta_c$
      and interaction strength $U/k_BT$ (polymer reservoir packing fraction
      $\eta_p^r$ and the reduced second virial coefficient $B_{2}^{*}$).
      The circles and the triangles indicate the stable fluid-solid coexistence region. (The circles are the results of  simulations where a total of $3 \times10^{4} $ MC moves per particles were used. The triangles are the results of more accurate simulations where  a total of $3\times10^{5} $ MC moves per particles were used).
      The continuous lines are a guide to the eye.
      The squares indicate the calculated coexistence points of the gas-liquid binodal. The thin dashed line is a fit to the binodal.
      The points indicate the state points where we carried out BD simulations.
      The red (dark grey) region indicates where the crystal fraction  is larger than 0.4 at $t=600 \tau_{B}$.
      The orange (light grey) region indicates where the crystal fraction  is smaller than 0.4 at $t=600 \tau_{B}$.
      }
   \label{fig:phdfull2}
\end{figure}

We determine the phase diagram by calculating free energies in
Monte Carlo simulations~\cite{Dijkstra1999a} for the potential (\ref{te:tot}) and for a size ratio
$q=0.15$, which corresponds approximately to the experimental
value of recent work~\cite{Lu2006}. Fig.~\ref{fig:phdfull2} shows
the equilibrium phase diagram  in the contact energy $\beta U$
(polymer reservoir packing fraction $\eta_{p}^{r}$), colloid
packing fraction $\eta_c$ representation. For $\beta U$=0 we
recover the hard-sphere phase diagram with a fluid phase at
$\eta_{c}\simeq 0.494$ in coexistence with a face-centered-cubic
(fcc) crystal phase at
 $\eta_{c}\simeq 0.545$. For increasing $\beta U$
the region of fluid-solid coexistence opens up. At the critical
energy $\beta U_{cr} \simeq$3.48 ( ($\eta_{p}^{r})_{cr} \simeq
0.316$) we find a gas-liquid transition which is metastable with respect to 
a broad fluid-solid coexistence.

\begin{figure}[htbp]
   \centering
   \includegraphics[width=8cm]{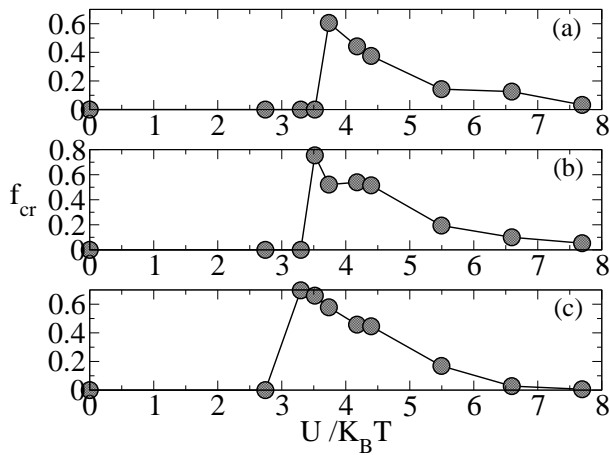}
      \caption{Crystal fraction $f_{\mbox{\small cr}}$ at $t=600 \tau_{B}$ as a function of the
      interaction energy $U/k_BT$ for (a) $\eta_{c}$=0.10. (b) $\eta_{c}$=0.25. (c) $\eta_{c}$=0.40.}
   \label{fig:fraccry}
\end{figure}

In order to follow the crystallization, we characterize the local
structure around particle $i$ by the $q_6$ order
parameter~\cite{Wolde1995}. We define the crystal fraction
$f_{cr}(t)$ as the number of crystal-like particles at time $t$
divided by the total number of particles. The red (dark grey) region in
Fig.~\ref{fig:phdfull2} indicates the statepoints for which the
crystal fraction is larger than 0.4 at $t=600 \tau_B$ while the orange (light grey) region indicates a fraction of crystal smaller than 0.4.
Let us discuss first the low $\beta U$ limit of the  red (dark grey) region.
Clearly, at low density the metastable gas-liquid binodal
has a dramatic effect in inducing (spontaneous) crystallization of
the colloidal particles, with the bottom boundary of the crystallization region as denoted by the red (dark grey) region in Fig.~\ref{fig:phdfull2} coinciding with the metastable gas-liquid binodal. 
Below the binodal line, and at low colloid density, we do not
observe spontaneous nucleation, in agreement with the observations
of~\cite{Charbonneau2007}. Neither do we observe, in the time scale of our simulations, crystallization
just below to the critical point despite the presence of critical
fluctuations that lower the nucleation barrier~\cite{Wolde1997}.
On the other hand, at medium-high colloid density we observe crystallization also outside the binodal. 
We now focus on the high $\beta U$ limit for crystallization. 
Fig.~\ref{fig:fraccry} shows that the crystal fraction slowly decays to zero
for increasing attractive energy $\beta U$,  and that the crystal fraction does not depend
strongly on $\eta_c$ for fixed $\beta U$. This indicates a
crystallization process hindered by a slowing down of the dynamics. The boundary between the orange (light grey) and red (dark grey) regions in Fig.~\ref{fig:phdfull2} is a horizontal line at $\beta U \approx 4.4$. As it has been demonstrated for short-ranged attractive systems, isodiffusivity lines are also horizontal
when approaching the gas-liquid binodal ~\cite{Foffi2005,Zaccarelli2007}.
Therefore, the slow decrease of crystallinity can be seen as an indicator of the proximity of the (attractive)
glass transition line that intersects the gas-liquid binodal. 
It is worthwile noting that the attractive glass transition was determined to be at $\eta_p^r \simeq 0.43$ using the long-time scaling of the incoherent correlation functions obtained from computer simulations for a system with a size ratio of q=0.1, and using a pair potential that included a long-range repulsive barrier to prevent gas-liquid phase separation~\cite{Puertas2005}. This value is remarkable close to our value of $eta_p^r \sim 0.4~$ for which we find hardly any crystallization. 

\begin{figure}[htbp]
   \centering
   \includegraphics[width=8cm]{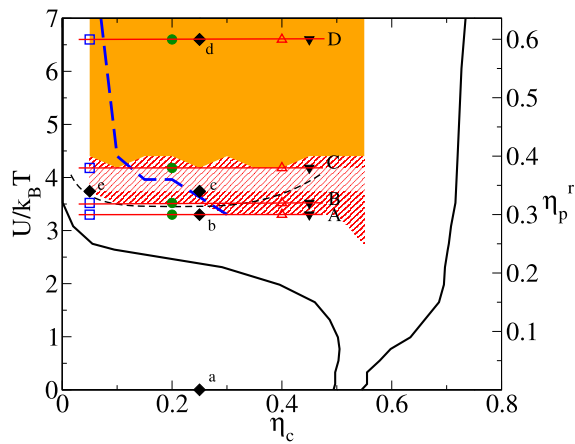}
      \caption{
      (Color online) The same as Fig.~\ref{fig:phdfull2}, but we indicate the state points discussed in the text. 
      For clarity the binodal symbols are removed . 
     The thick (blue) dashed line indicates the transition between percolating and non-percolating states.}
   \label{fig:phdfull3}
\end{figure}

\begin{figure}[htbp]
   \centering
     \includegraphics[width=8cm]{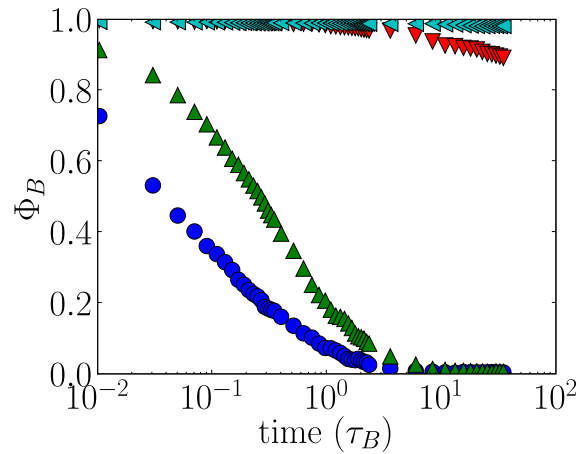}
 \caption{(Color online) Bond correlation functions $\Phi_{B}(t)$ as a function of time $t$ for  $\eta_{c}$=0.25.
      Circles indicate results for energy $\beta U=0$ (point a in Fig.~\ref{fig:phdfull3}), triangles-up indicate $\beta U=3.3$ (point b in Fig.~\ref{fig:phdfull3}), triangles-down indicate $\beta U=3.74$ (point c in Fig.~\ref{fig:phdfull3}), triangle-left indicate $\beta U=6.6$ (point d in Fig.~\ref{fig:phdfull3}).  }
   \label{fig:corr}
\end{figure}

Fig.~\ref{fig:phdfull3} shows the phase diagram as in Fig.~\ref{fig:phdfull2} but we label state points that will be discussed below. 
To demonstrate the existence of a dynamical arrest we compute the bond correlation function $\Phi_B(t)$ (Fig.~\ref{fig:corr}) measured starting at a time t=600 $\tau_{B}$ after the initial quench in the state points a-d in Fig.~\ref{fig:phdfull3}.
 We observe a divergence in the characteristic decay time between $\beta U=3.3$ (point b in Fig.~\ref{fig:phdfull3}) and $\beta U=3.74$ (point d in Fig.~\ref{fig:phdfull3}), giving evidence of a 
fluid-solid (either crystalline or amorphous) transition. 
The behavior of $\Phi_B(t)$ is very similar whether the solidification is due to crystallization ($\beta U=3.74$) or 
an attractive glass transition ($\beta U=6.6$).

We now focus our attention to the state points inside the
gas-liquid region. We find that in certain regions of the phase
diagram the phase separation is
arrested in a permanently percolated structure. The thick dashed line in
Fig.~\ref{fig:phdfull3} corresponds to the percolation limit. 
For quenches inside the gas-liquid coexistence we find, on the left side of the percolation line, clusters that are
crystalline at low $\beta U$ (red (dashed) region in Fig.~\ref{fig:phdfull3}) and amorphous at high $\beta U$ (orange (light grey) region in Fig.~\ref{fig:phdfull3}), while on the right-hand side, 
we find a  `long-lived' percolating
network (gel) with high-density branches that are crystalline at low
$\beta U$ (red (dashed) region in Fig.~\ref{fig:phdfull3}) and amorphous at high $\beta U$ (orange (light grey) region in Fig.~\ref{fig:phdfull3}).

\begin{figure}[htbp]
   \centering
   \includegraphics[width=8cm]{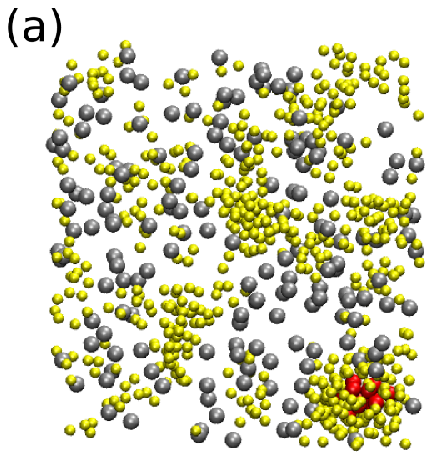}
     \includegraphics[width=8cm]{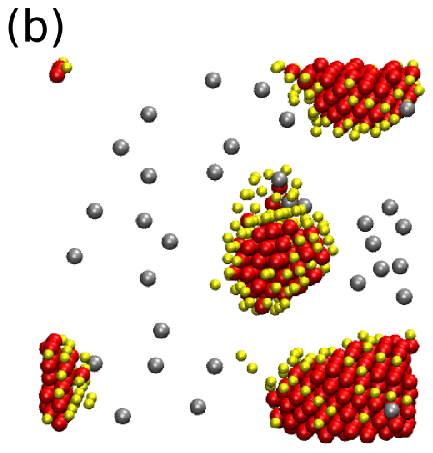}
 \caption{ (Color online) Typical snapshots for the state point e in Fig.~\ref{fig:phdfull3} ($\eta_{c}$=0.05 and $\beta U=3.74$) at  t=50 $\tau_{B}$ (a), and at t=600 $\tau_{B}$ (b).
 Gas particles are indicated by large light grey spheres, fluid particles are indicated by yellow
 (small light grey) spheres and crystal particles are indicated by blue (large dark grey) spheres.
 The complete kinetic pathway is shown in the attached movies~\cite{movies}. }
 \label{fig:path1}
\end{figure}

We now analyze in more detail the kinetic pathways in different
regions of the phase diagram. The first state point we consider is point e in Fig.~\ref{fig:phdfull3} (
$\eta_{c}$=0.05, $\beta U = 3.74$, and $\eta_{p}^{r}$=0.34), just
beyond the gas binodal. The direct path to the equilibrium phase
would be nucleation of the crystal phase in the dilute fluid
phase. The actual kinetic pathway is shown in the attached movies~\cite{movies}. 
Fig.~\ref{fig:path1}(a) and (b) show typical
configurations at $t=50 \tau_{B}$ and $t=600 \tau_{B}$,
respectively. We observe first in Fig.~\ref{fig:path1}(a) the
formation of metastable liquid droplets (yellow/small light
spheres)  surrounded by gas particles (large light spheres).
Subsequently the nucleation of the crystal (red/large dark 
spheres)  starts within the fluid droplets. The crystalline
clusters grow until a local equilibrium between the cluster and a
surrounding gas phase is reached. These metastable clusters
survive for a long time (Fig.~\ref{fig:path1}(b)) due to low
diffusion, but irreversibly coalesce upon touching, in contrast
with the findings of~\cite{Lu2006}.
\begin{figure}[htbp]
   \centering
       \includegraphics[width=8cm]{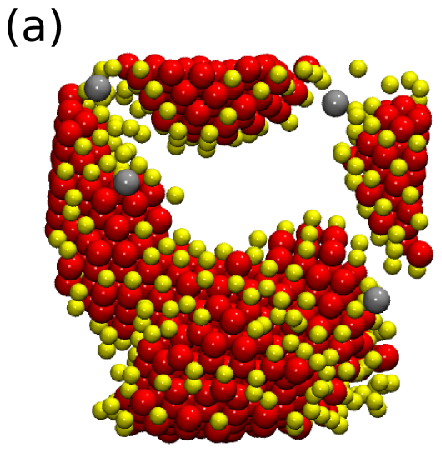}
         \includegraphics[width=8cm]{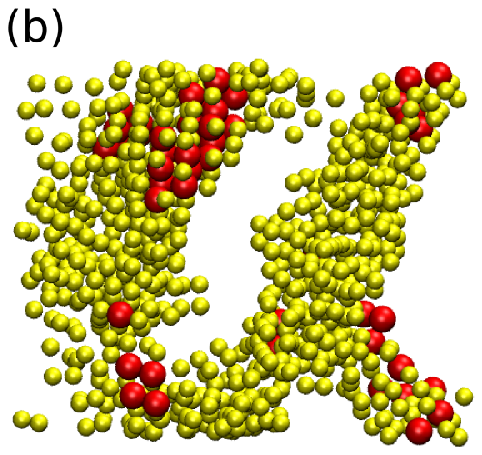}
 \caption{ (Color online) A snapshot for state point c in Fig.~\ref{fig:phdfull3} ($\eta_{c}$=0.25 and $\beta U=3.74$) at t=600 $\tau_{B}$) (a) and for state point d in Fig.~\ref{fig:phdfull3} ($\eta_{c}$=0.25 and $\beta U=6.6$) at t=600 $\tau_{B}$ (b). Gas particles are indicated by large light grey spheres, fluid particles are indicated by yellow
 (small light grey) spheres and crystal particles are indicated by blue (large dark grey) spheres. The complete kinetic pathway is shown in the attached movies~\cite{movies}.}
 \label{fig:path2}
\end{figure}
At higher colloid density $\eta_c=0.25$ and $\beta U$=3.74 (point c in Fig.~\ref{fig:phdfull3}), we
observe spinodal decomposition, and subsequently nucleation of
crystal nuclei within the fluid branches of the spinodal
structure. The crystalline clusters grow until the
whole spinodal structure is crystalline. Fig.~\ref{fig:path2}(a)
shows that the spinodal structure is arrested by crystallization.
At higher interaction strength $\beta U$=6.6 and $\eta_{c}$=0.25 (point d in Fig.~\ref{fig:phdfull3}),
we observe spinodal decomposition, and only very small crystal
nuclei appear within the fluid branches of the spinodal structure.
The spinodal decomposition is arrested into a branched gel by the
formation of  a glassy solid (Fig.~\ref{fig:path2}(b)). This
scenario is consistent with non-equilibrium gelation induced by
spinodal decomposition arrested by an attractive glass transition.
The metastable gas-liquid phase separation is arrested due to
decreased diffusion of  single particles by crystallization at low $\beta U$.
On the other hand, at high $\beta U$ the phase separation is arrested by a slow dynamics that is consistent with vitrification, but we do not explicitly calculate the exact location of the glass transition here. 

\begin{figure}[htbp]
   \centering
   \includegraphics[width=7cm]{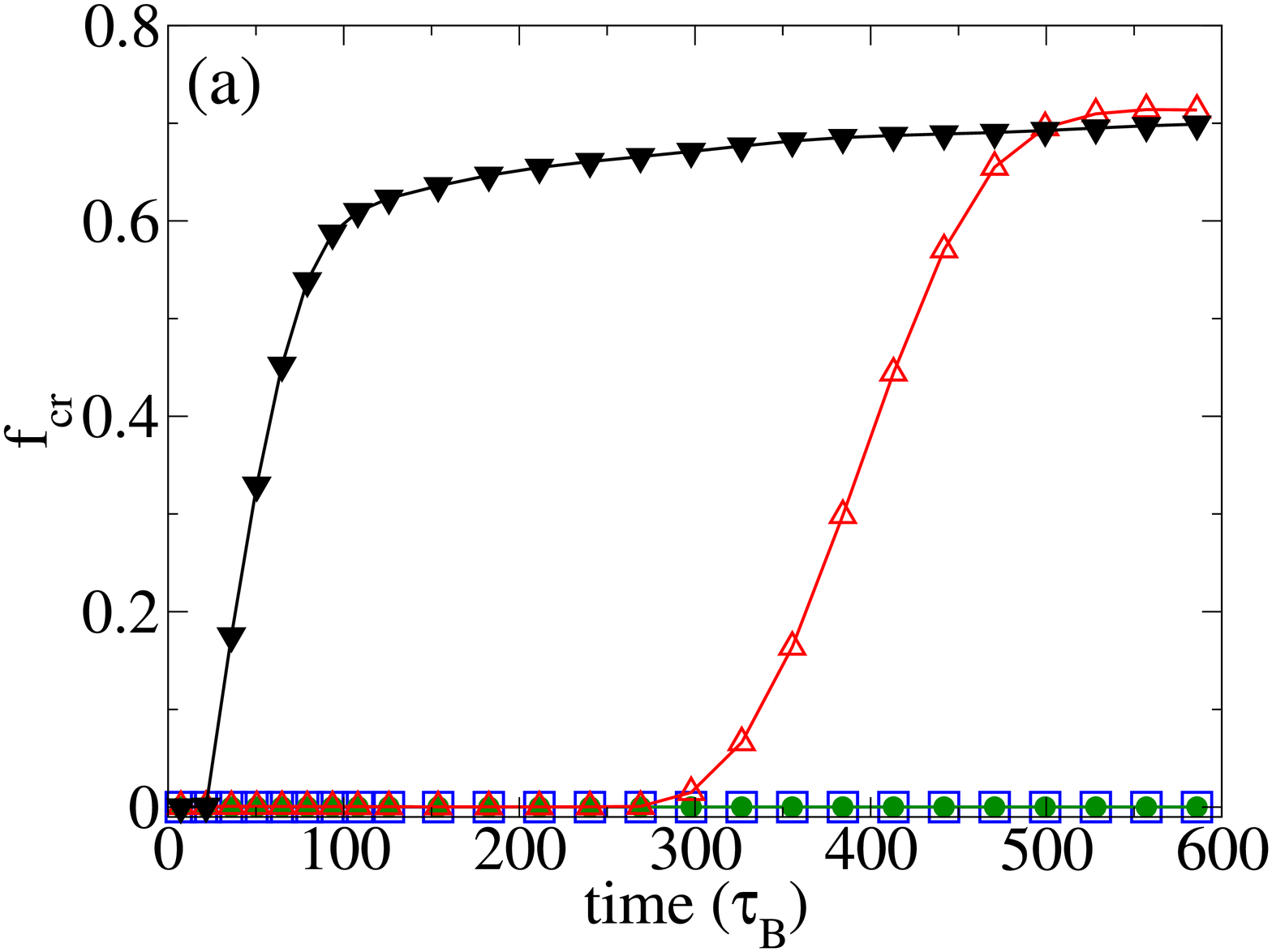}
    \includegraphics[width=7cm]{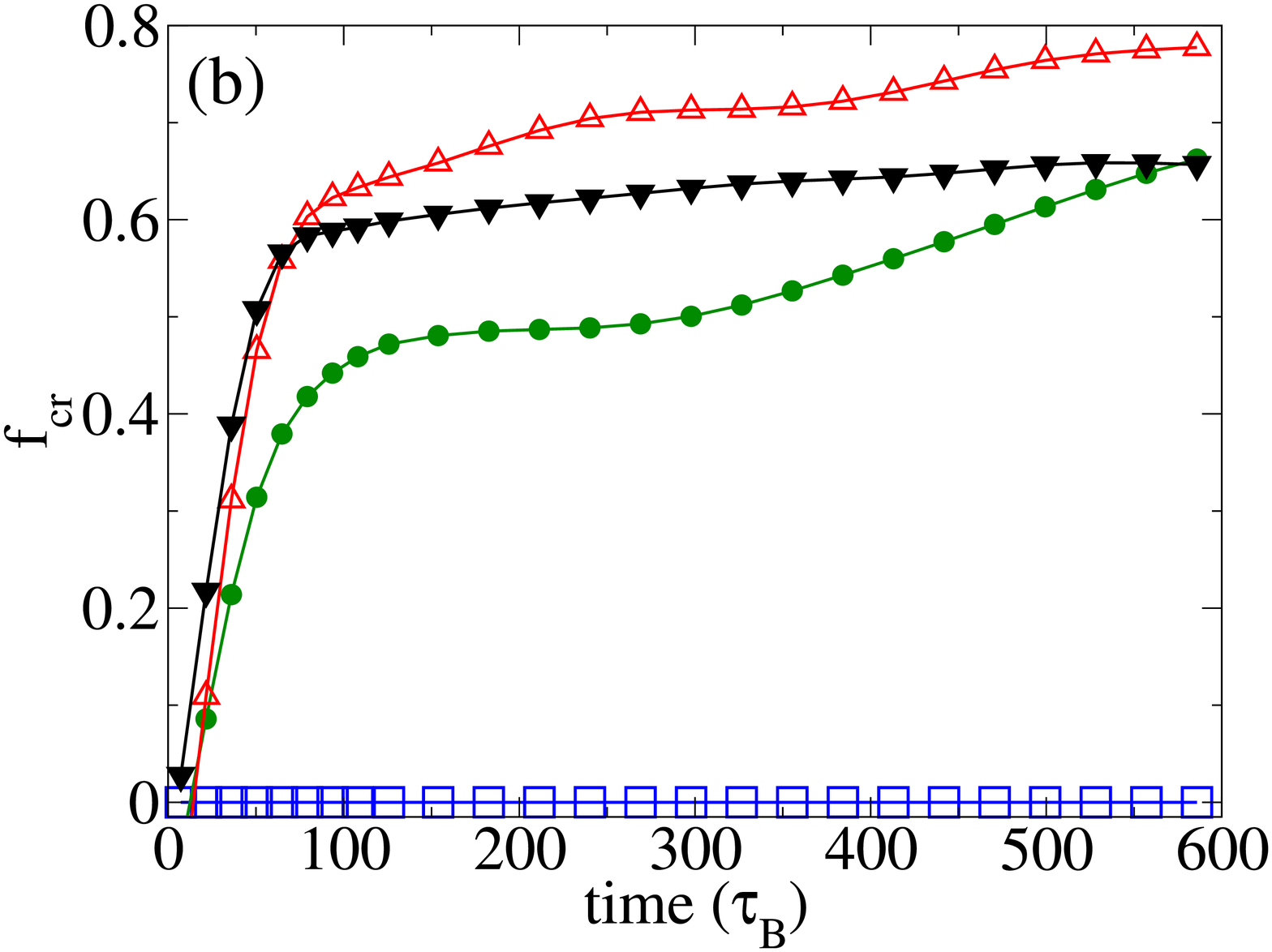}
     \includegraphics[width=7cm]{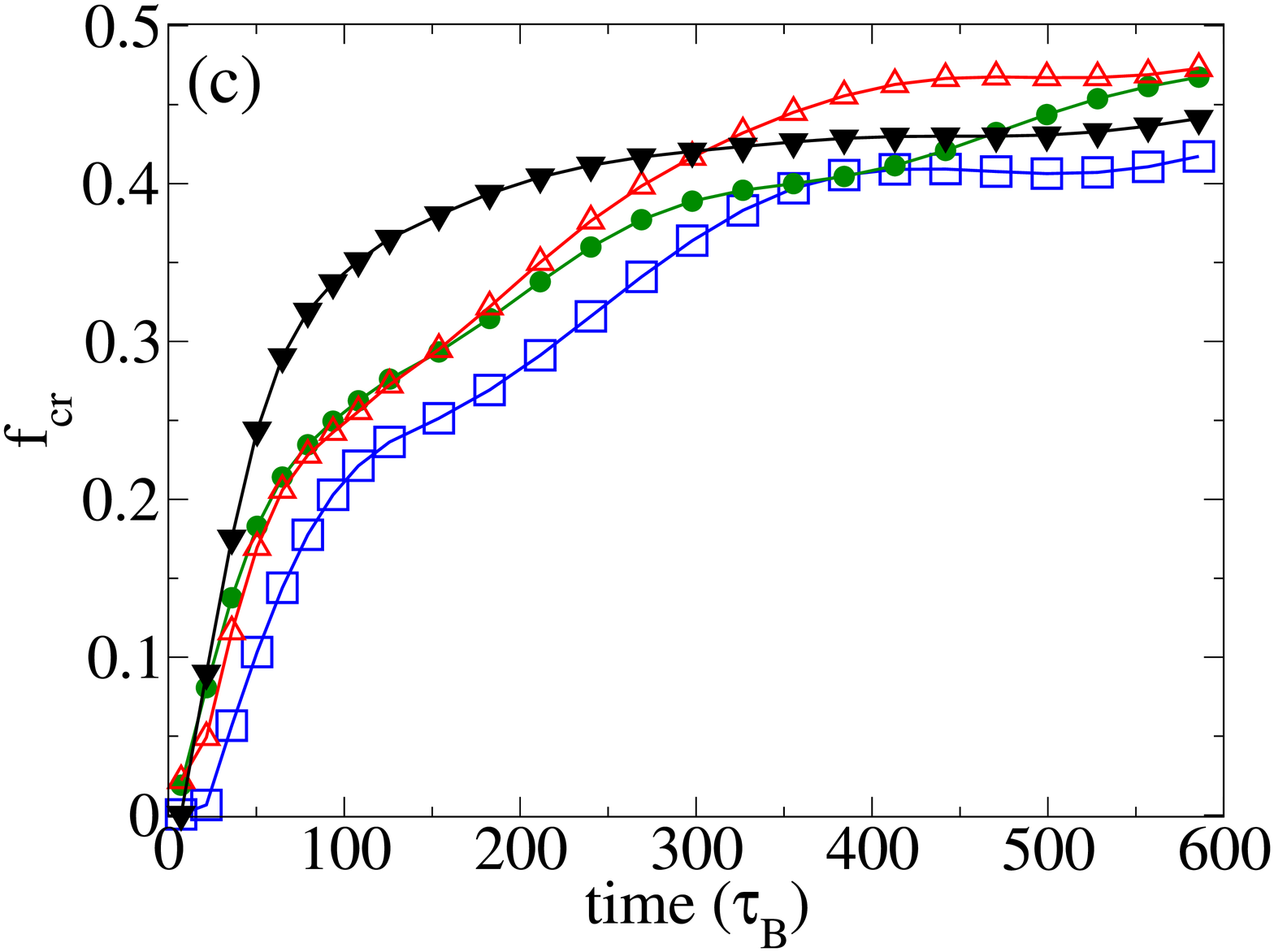}
      \includegraphics[width=7cm]{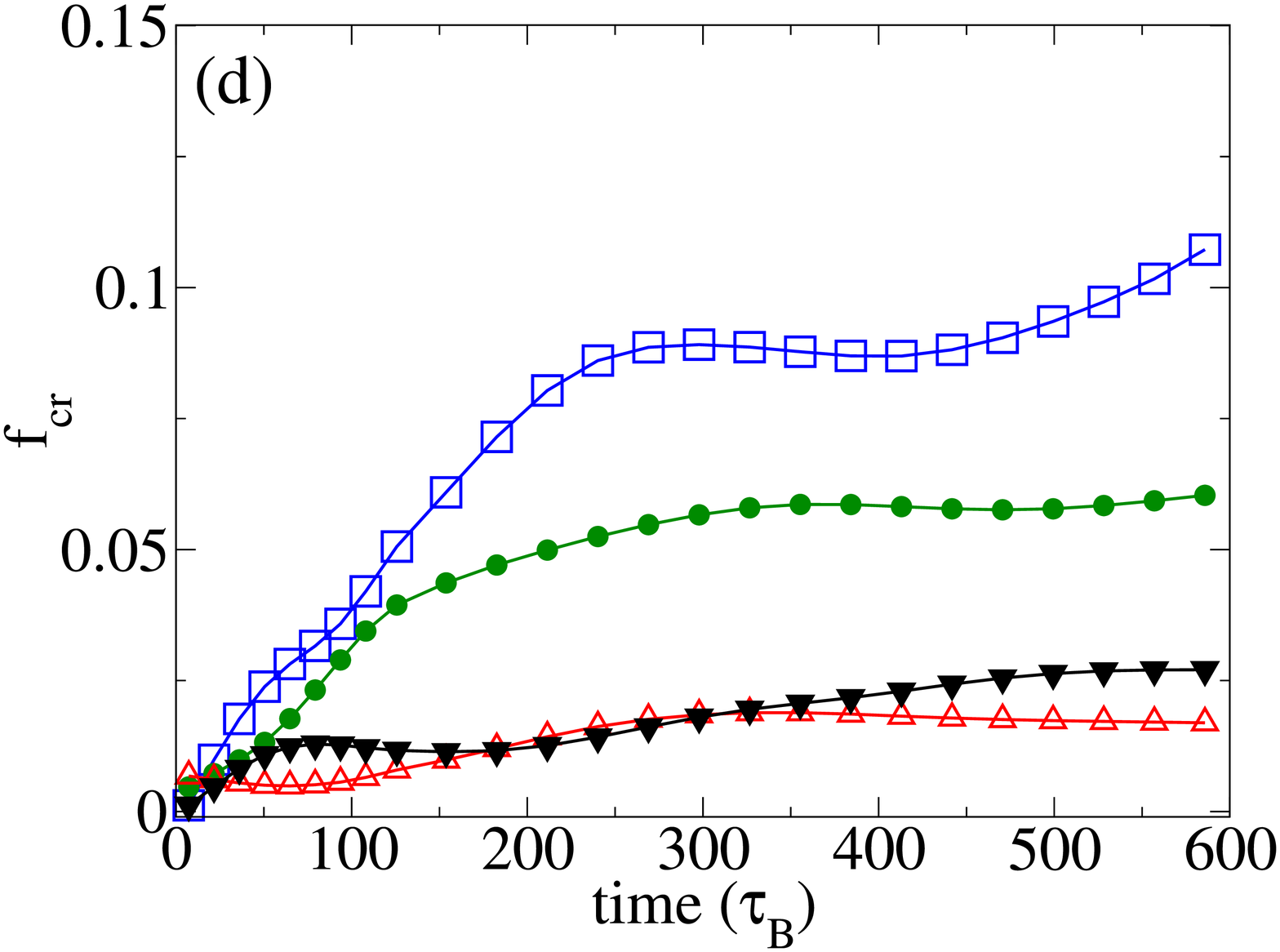}
      \caption{ (Color online) Crystal fraction $f_{\mbox{cr}}$ as a function of time $t$ for (a) $\beta U$=3.3 (line A in Fig.~\ref{fig:phdfull3}), (b) $\beta U$=3.52 (line in B Fig.~\ref{fig:phdfull3}), (c) $\beta U$=4.18 (line in C in Fig.~\ref{fig:phdfull3}), (d) $\beta U$=6.6 (line D Fig.~\ref{fig:phdfull3}). 
      Squares indicate a colloid packing fraction $\eta_{c}$=0.05, circles indicate $\eta_{c}$=0.20, triangles-up indicate $\eta_{c}$=0.40, and triangles-down indicate $\eta_{c}$=0.45
 (same symbols as in Fig. 3)}
   \label{fig:growth}
\end{figure}

Fig.~\ref{fig:growth} shows the crystal fraction
$f_{\mbox{cr}}(t)$ as a function of time $t$  for $\beta U$=3.3,
3.52, 4.18, and 6.6  and for varying $\eta_{c}$ (filled symbols along the red lines A-D in Fig.~\ref{fig:phdfull3}). 
The evolution of the crystal fraction is characterized by three distinct regimes.
In the first regime the fraction of crystal is equal to zero. This regime corresponds to the time elapsed between the quench and the formation of a critical nucleus (induction time). The second regime corresponds to the fast growth of the  crystal.
Fig.~\ref{fig:growth}(a) shows that for statepoints $\eta_c$=0.05, 0.20, and 0.40 of line A in Fig.~\ref{fig:phdfull3}, 
the induction time is significantly different from zero. We note here that the induction time is larger than 600 $\tau_B$ for $\eta_c=0.05$ and 0.20. All these statepoints are outside the metastable gas-liquid binodal region. 
In Fig.~\ref{fig:growth}(b), we find that statepoint $\eta_c=0.05$ of line B in Fig.~\ref{fig:phdfull3} shows no crystallization at all, and hence the induction time is again larger than $600 \tau_B$. 
This statepoint also lies outside the gas-liquid coexistence region, i.e, $\eta_c<\eta_c^{sat}$, where $\eta_c^{sat}$ denotes the coexisting packing fraction of the metastable gas phase. For quenches to $\eta_c>\eta_c^{sat}$
, i.e., $\eta_c$=0.20, 0.40 and 0.45, we find that the induction time is close to zero.  Similarly, for quenches inside the gas-liquid spinodal as shown in Fig.~\ref{fig:growth}(c) and \ref{fig:growth}(d) corresponding to line C and D in Fig.~\ref{fig:phdfull3}, the induction time is also close to zero as the formation of the spinodal structure is followed immediately by nucleation of small crystallites and subsequent growth of the crystals. In this case, the induction time is related to the spinodal decomposition $\simeq \tau_D$~\cite{Sanz:2008} that is not visible on the timescale of figure~\ref{fig:growth}. To summarize, we find that the induction time is close to zero for all quenches beyond the gas binodal of the metastable gas-liquid coexistence region, i.e., for $\eta_c>\eta_c^{sat}$. 
The fraction of crystal grows rapidly after the induction time due to the huge difference in chemical potential between the liquid and the growing crystal phase. 
In the third regime, the crystal fraction grows slowly by coarsening and defect removal processes.
The fraction of crystal reached by the system depends strongly on the contact energy. At low $\beta U$ particle diffusion is fast enough for crystallization. On the other hand, at high $\beta U$ the particles diffuse slowly and despite the thermodynamic drive to form the crystal the system is trapped in an amorphous solid. 
To summarize, for state points inside the binodal we always observe an induction time equal to zero, consistent with the 
nucleation of small crystallites in the initial stage, and subsequent growth of the crystals. 
Outside the binodal, the induction time is close to zero only at high colloid packing fraction.
Furthermore, we observe that the fraction of crystal is independent of the packing fraction for fixed $\beta U$, while it decreases for increasing contact energy $\beta U$.

\begin{figure}[htbp]
   \centering
     \includegraphics[width=8cm]{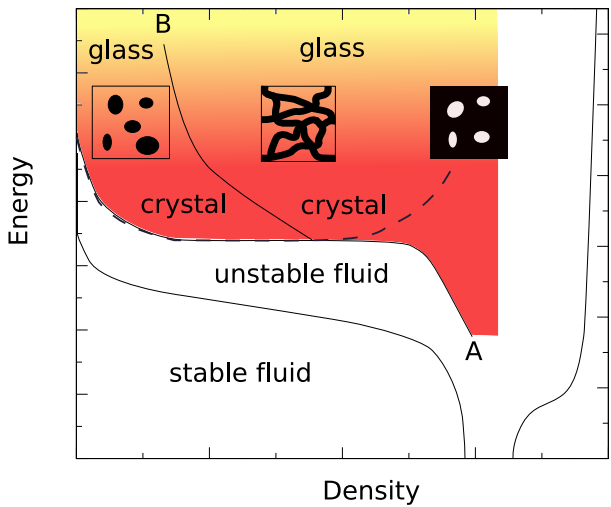}
      \caption{ (Color online) Sketch of the phase diagram in the energy-density representation reporting the location of the non-equilibrium phases found in this work. Line A is the line of spontaneous crystallization while Line B indicates the transition form a cluster phase to a gel. 
 The square insets sketch the mesoscopic morphology of the system
(clusters, percolating network (gel) and porous solid from left to right).The shading indicate the transition from a more crystalline phase (dark red) to a more amorphous one (light yellow).}
   \label{fig:sum}
\end{figure}
\section{Conclusions}
The objective of this work was to systematically  study the
relationship between equilibrium and non-equilibrium phases in
systems with short-ranged attractions. To study the dynamics of
the system we carried out BD simulations  by quenching a homogeneous fluid phase to certain state points in the  phase
diagram. 

First, we find a window of enhanced crystallization that is delimited at high contact energy by low diffusion and at low contact energy 
 (below line A Fig.~\ref{fig:sum}) by a high nucleation barrier.

Second, we observe that the crystallization is enhanced by the
metastable gas-liquid phase separation, as demonstrated by the fact that the line of spontaneous crystallization (line A in Fig.~\ref{fig:sum}) at low density 
follows closely the binodal of the gas-liquid
separation (dashed line in Fig.~\ref{fig:sum}).  The effect is caused by the two-step process crystallization process inside the gas-liquid binodal:
the system first forms liquid domains, either through nucleation or spinodal decomposition; subsequently, crystal nucleation occurs within the 
dense liquid domains.
This process follows the Ostwald rule of stages that states that
the transition from one phase to another  may proceed via
intermediate metastable states.  Hence, the effective
nucleation rate of the crystal depends on the nucleation rate of
the liquid in the gas and the crystal in the liquid phase. This two-stage crystallization
process was also observed  in
proteins~\cite{Lomakin2003,Filobelo2005}.

Last, we find two competing mechanisms of arrest of the gas-liquid spinodal decomposition.
At low $\beta U$, we observe that
crystallization starts within the liquid branches before the
spinodal structure can coarsen into macroscopic phase separation.
The crystallization slows down the coarsening process and  arrests
the system in structures typical of spinodal decomposition. For
higher $\beta U$, the system becomes  amorphous as crystallization
is inhibited. This result supports a gelation caused by a glass transition that meets the liquid binodal and 
thereby arrests the gas-liquid phase separation~\cite{Lu:2008}. 
We find that crystallization and vitrification are both responsible
for the formation of arrested spinodal patterns (gels), albeit at different interaction
energies.
These two mechanisms are also responsible for the formation of two different types of long-lived, but transient clusters at low colloid density.
The clusters are crystalline at low $\beta U$ and glassy at high $\beta U$.
At high colloid density, on the right hand side of the gas-liquid binodal, the amorphous arrested phase is
structurally very similar to a (porous) attractive glass, while the
crystalline arrested phase can be seen as a polycrystalline porous
material.   Fig.~\ref{fig:sum} shows schematically the location of the different type of phases we found in relation to the equilibrium phase diagram.

Finally, we point out that our phase diagram can provide a guide to experimentally obtain and control the morphology of crystals of colloids or proteins. 
The crystallization window can be located experimentally since it is delimited at low temperature by an arrested phase with liquid-like local 
structure and at high temperature by a mobile fluid phase. 
Thus, by changing the temperature it is possible to control the degree of crystallinity. 
The highest fraction of crystal is obtained for state points in the lower interaction strength (higher temperature) 
part of the crystallization window. For increasing  interaction strength (decreasing temperature) the crystal fraction decreases.     
For the range of packing fractions studied in this work ($\eta_{c}< 0.55$), the colloid packing fraction does not influence the fraction of crystal or the growth rate, but has a strong influence on the morphology of the final phase.
While at very low $\eta_{c}$ clusters form,  a branched structure is obtained at intermediate packing fractions, and porous materials are obtained at high $\eta_{c}$. 

\acknowledgments
We thanks Neer Asherie for critically reading the manuscript. 
This work is part of the research
program of the {\em Stichting voor Fundamenteel Onderzoek der
Materie} (FOM), that is financially supported by the {\em
Nederlandse Organisatie voor Wetenschappelijk Onderzoek} (NWO). 
NWO-CW is acknowledge for the TOP-CW funding.


\end{document}